\documentclass[preprintnumbers,article,amsmath,amssymb,floatfix,10pt,prd,twocolumn,
superscriptaddress,nofootinbib]{revtex4-2}
\usepackage{bm}
\usepackage{amsfonts}
\usepackage{latexsym}
\usepackage[latin1]{inputenc}
\usepackage{graphicx}
\usepackage{amsmath}
\usepackage{palatino}
\usepackage{mathpazo}
\usepackage{textcomp}
\usepackage{float}
\usepackage{booktabs}
\usepackage{dcolumn}
\usepackage{ragged2e}
\usepackage{hyperref}
\hypersetup{colorlinks,citecolor=blue}
\hypersetup{colorlinks=true,linkcolor=red,filecolor=magenta,    urlcolor=blue}
\usepackage{amsmath}
\usepackage{xcolor}
\usepackage{orcidlink}
\usepackage{epsfig}
\usepackage{caption}
\usepackage{subcaption}
\usepackage{commath}
\def\jnl@style{\it}
\def\aaref@jnl#1{{\jnl@style#1}}

\def\aaref@jnl#1{{\jnl@style#1}}

\def\aj{\aaref@jnl{AJ}}                   
\def\apj{\aaref@jnl{ApJ}}                 
\def\apjl{\aaref@jnl{ApJ}}                
\def\apjs{\aaref@jnl{ApJS}}               
\def\apss{\aaref@jnl{Ap\&SS}}             
\def\aap{\aaref@jnl{A\&A}}                
\def\aapr{\aaref@jnl{A\&A~Rev.}}          
\def\aaps{\aaref@jnl{A\&AS}}              
\def\mnras{\aaref@jnl{Mon.~Not.~Roy.~Astron.~Soc.}}             
\def\prd{\aaref@jnl{Phys.~Rev.~D}}        
\def\prc{\aaref@jnl{Phys.~Rev.~C}}  
\def\prl{\aaref@jnl{Phys.~Rev.~Lett.}}    
\def\qjras{\aaref@jnl{QJRAS}}             
\def\skytel{\aaref@jnl{S\&T}}             
\def\ssr{\aaref@jnl{Space~Sci.~Rev.}}     
\def\zap{\aaref@jnl{ZAp}}                 
\def\nat{\aaref@jnl{Nature}}              
\def\aplett{\aaref@jnl{Astrophys.~Lett.}} 
\def\apspr{\aaref@jnl{Astrophys.~Space~Phys.~Res.}} 
\def\physrep{\aaref@jnl{Phys.~Rep.}}      
\def\physscr{\aaref@jnl{Phys.~Scr}}       
\def\commat{\aaref@jnl{Comm.~Math.~Phys.}}              
\def\science{\aaref@jnl{Science}}               
\def\cqg{\aaref@jnl{Classical Quant.~Grav.}}            
\def\jpcs{\aaref@jnl{JPCS}}                                     
\def\ijmpd{\aaref@jnl{Int.~J.~Mod.~Phys.~D}}                    
\def\grg{\aaref@jnl{Gen.~Relat.~Gravit.}}               
\def\rpp{\aaref@jnl{Rep.~Prog.~Phys.}}          
\def\npa{\aaref@jnl{Nucl.~Phys.~A}}        
\def\lrr{\aaref@jnl{Living Rev.~Rel.}}                   
\def\jcap{\aaref@jnl{J.~Cosmology Astropart.~Phys.}}    
\def\rmp{\aaref@jnl{Rev.~Mod.~Phys.}}   
\def\epjc{\aaref@jnl{Eur.~Phys.~J.~C}} 
\def\plb{\aaref@jnl{~Phy.~Lett.~B}} 
\def\mpla{\aaref@jnl{Mod.~Phy.~Lett.~A}} 
\def\arxiv{\aaref@jnl{arxiv.org}}

\allowdisplaybreaks[1]

\begin{document}
\color{black}       
\title{Bulk viscous matter in $f(T)$ gravity: A path to cosmic acceleration}

\author{Kairat Myrzakulov\orcidlink{0000-0002-4189-8596}}\email[Email: ]{krmyrzakulov@gmail.com} 
\affiliation{Department of General \& Theoretical Physics, L.N. Gumilyov Eurasian National University, Astana, 010008, Kazakhstan.}

\author{O. Donmez\orcidlink{0000-0001-9017-2452}}
\email[Email: ]{orhan.donmez@aum.edu.kw}
\affiliation{College of Engineering and Technology, American University of the Middle East, Egaila 54200, Kuwait.}

\author{M. Koussour\orcidlink{0000-0002-4188-0572}}
\email[Email: ]{pr.mouhssine@gmail.com}
\affiliation{Department of Physics, University of Hassan II Casablanca, Morocco.}

\author{S. Muminov\orcidlink{0000-0003-2471-4836}}
\email[Email: ]{sokhibjan.muminov@gmail.com}
\affiliation{Mamun University, Bolkhovuz Street 2, Khiva 220900, Uzbekistan.}

\author{E. Davletov\orcidlink{0000-0003-1132-8178}}
\email[Email: ]{davletov_erkaboy@mamunedu.uz}
\affiliation{Mamun University, Bolkhovuz Street 2, Khiva 220900, Uzbekistan.}

\author{J. Rayimbaev\orcidlink{0000-0001-9293-1838}}
\email[Email: ]{javlon@astrin.uz}
\affiliation{Institute of Fundamental and Applied Research, National Research University TIIAME, Kori Niyoziy 39, Tashkent 100000, Uzbekistan.}
\affiliation{University of Tashkent for Applied Sciences, Str. Gavhar 1, Tashkent 100149, Uzbekistan.}
\affiliation{Urgench State University, Kh. Alimjan Str. 14, Urgench 221100, Uzbekistan}
\affiliation{Shahrisabz State Pedagogical Institute, Shahrisabz Str. 10, Shahrisabz 181301, Uzbekistan.}


\begin{abstract}

In this paper, we investigate the effects of varying bulk viscosity coefficients $\zeta(t)=\zeta_{0}+\zeta_{1}H$ on cosmic evolution within the framework of $f(T)$ teleparallel gravity. We focus on two cases: (i) $\zeta_{1} \neq0$ and (ii) $\zeta_{1} =0$, deriving the Hubble parameter $H$ as a function of redshift $z$ using a linear $f(T)$ model ($f(T) = \alpha T$ where $\alpha \neq 0$). Using the combined $H(z)+Pantheon^{+}+BAO$ dataset, we obtain observational constraints on model parameters. For Case I ($\zeta_1 \neq 0$), best-fit values are $H_0=60.0^{+2.0}_{-1.9}$ km/s/Mpc, $\alpha=1.01^{+0.10}_{-0.098}$, $\zeta_0=40.1^{+1.9}_{-2.0}$, and $\zeta_1=0.123^{+0.093}_{-0.088}$, while for Case II ($\zeta_1 = 0$), they are $H_0=67.5^{+1.3}_{-1.3}$ km/s/Mpc, $\alpha=0.94^{+0.14}_{-0.13}$, and $\zeta_0=34.7^{+2.0}_{-2.0}$. The analysis reveals a transition in the deceleration parameter, indicating a shift from deceleration to acceleration of the universe's expansion, with present-day values of $q_{0} \approx -0.49$ and $q_{0} \approx -0.32$ for the respective cases. The jerk parameter $j(z)$ and effective EoS for the cosmic viscous fluid also support the cosmic acceleration, with trajectories aligning with the quintessence scenario. These findings underscore the potential of our $f(T)$ model dominated by bulk viscous matter in explaining cosmic acceleration.\\

\textbf{Keywords:} bulk viscosity, $f(T)$ teleparallel gravity, statefinder parameters, and $Om(z)$ diagnostic.

\end{abstract}

\maketitle

\section{Introduction}

Presently, general relativity (GR) stands as the cornerstone of modern gravitational physics, acclaimed for its unparalleled success in describing the gravitational interaction on cosmic scales. Einstein's revolutionary theory, published in 1915, fundamentally transformed our understanding of space, time, and gravity \cite{Hilbert/1915,Einstein/1915,Einstein/1916}. The theory's striking predictions, including the perihelion advance of Mercury, the deflection of light by the Sun, gravitational redshift \cite{Landau/1970}, and radar echo delay \cite{Shapiro/1971,Shapiro/1976}, have been confirmed with unmatched accuracy through observational evidence. Furthermore, forecasts like the orbital decay observed in the Hulse-Taylor binary pulsar, attributed to gravitational-wave damping, have also substantiated the theory's weak-field validity through observations \cite{Taylor/1982}. For comprehensive reviews of experimental and observational tests of GR, refer to \cite{Will/2014,Ishak/2019}. The detection of gravitational waves \cite{Abbott/2016} provided an opportunity to assess GR's predictions during the late stages of binary black hole coalescence, representing a scenario with strong gravitational fields.

Recent observational advancements in cosmology, including Type Ia Supernovae (SNe Ia) \cite{Riess/1998,Perl/1999}, Baryon Acoustic Oscillations (BAOs) \cite{Eisenstein/2005, Percival/2007}, and the Cosmic Microwave Background (CMB) \cite{Komatsu/2011}, have compellingly demonstrated that our universe has recently entered a phase of accelerated expansion. Furthermore, these observations suggest a surprising outcome: approximately 95\% of the universe's content consists of two enigmatic components known as dark energy (DE) and dark matter (DM), respectively, while only 5\% of the overall composition is made up of baryonic matter (BM). These observations have highlighted the shortcomings of GR. Despite its significant accomplishments and remarkable success at the scale of the solar system, it may not be sufficient to comprehensively explain gravitational phenomena on galactic and cosmological scales. Therefore, GR might not serve as the ultimate theory of the gravitational force. It fails to provide satisfactory explanations for the two fundamental problems confronting present-day cosmology: the DM problem and the DE problem.

Various approaches have been proposed recently to explain the observational results of cosmology. However, a satisfactory theory of gravity remains elusive. One possibility for developing new gravitational theories is to consider that the Einstein gravity model of GR breaks down at large scales, necessitating a more general action than the Hilbert-Einstein action. An elementary approach to extending Einstein's gravity is through $f(R)$ modified gravity, where an arbitrary function $f$ of the curvature scalar $R$ is introduced into the gravitational action \cite{Buchdahl/1970,Barrow/1983,Staro/2007,Capo/2008,Chiba/2007,Astorga/2023}. Other theories have emerged, such as $f(R,\mathcal{T})$ gravity, where $\mathcal{T}$ represents the trace of the energy-momentum tensor \cite{Harko/2011,Moraes/2017,Amani/2015}, $f(Q)$ gravity, where $Q$ represents the non-metricity scalar \cite{Jim/2018}, and $f(R,G)$ gravity, where $G$ represents the Gauss-Bonnet scalar \cite{Laurentis/2015,Gomez/2012}. However, a significant advancement in geometry, leading to a new class of generalized geometric theories of gravity, was made by Weitzenb\"ock, who introduced what are now known as Weitzenb\"ock spaces \cite{Weitz/1923}. A Weitzenb\"ock manifold is characterized by the non-metricity scalar vanishing ($Q=0$), the torsion scalar not vanishing ($T \neq 0$), and the curvature scalar vanishing ($R=0$) for the manifold. The Weitzenb\"ock manifold transforms to a Euclidean manifold at $T = 0$. Different regions of the Weitzenb\"ock manifold have different values for the torsion tensor. Because the curvature scalar of a Weitzenb\"ock space is zero, these geometries possess the significant feature of distant parallelism, also referred to as absolute parallelism or teleparallelism. Einstein was the first to apply Weitzenb\"ock-type spacetimes in physics, proposing a unified teleparallel theory that encompassed electromagnetism and gravity \cite{Einstein/1928}. In the teleparallel gravity, the fundamental concept involves replacing the spacetime metric $g_{\mu\nu}$, which is the primary physical quantity describing gravitational properties, with a set of tetrad vectors $e^{i}_{\mu}$. The torsion, arising from the tetrad fields, can be employed to fully characterize gravitational phenomena, replacing the role of curvature. This leads us to the teleparallel equivalent of GR (TEGR), originally formulated in \cite{Moller/1961,Pellegrini/1963,Hayashi/1979}, and currently recognized as the $f(T)$ gravity theory. Therefore, in $f(T)$ teleparallel gravity, torsion precisely counteracts curvature, leading to the spacetime becoming flat. A key benefit of the $f(T)$ gravity is that its field equations are second-order, contrasting with $f(R)$ gravity, which, in the metric approach, is a fourth-order theory. For an in-depth exploration of teleparallel theories, refer to \cite{Aldrovandi/2013}. Several studies in $f(T)$ gravity have investigated various aspects such as cosmological solutions \cite{Paliathanasis/2016}, thermodynamics \cite{Salako/2013}, late-time acceleration \cite{Myrzakulov/2011, Bamba/2011}, cosmological perturbations \cite{Chen/2011}, large-scale structure \cite{Li/2011}, cosmography \cite{Capozziello/2011}, energy conditions \cite{Liu/2012}, matter bounce cosmology \cite{Cai/2011}, wormholes \cite{Jamil/2013}, anisotropic universe \cite{Rodrigues/2016,Koussour/2022}, observational constraints \cite{Nunes/2016}, and viscous cosmology \cite{Yang/2022}. Recently, Ganjizadeh et al. \cite{Ganjizadeh/2022} analyzed observational Hubble parameter data to constrain an interactive gravity model with particle creation, examining the role of particle production in accelerating the universe's expansion. Rezaei and Amani \cite{Rezaei/2017} investigated the stability of extended $f(T)$ gravity with an energy-momentum tensor coupling in the context of a modified Chaplygin gas.

In the majority of cosmological models, the universe's contents are typically treated as a perfect fluid. It is crucial to explore more realistic models that incorporate dissipative processes arising from viscosity. In a homogeneous and isotropic universe, bulk viscosity is the sole viscous effect capable of altering the background dynamics. It is widely recognized that during neutrino decoupling, matter exhibited characteristics of a viscous fluid in the early stages of the universe \cite{Misner/1968,Israel/1970,Murphy/1973,Belinskii/1974}. In the inflation framework, it has long been recognized that an imperfect fluid with bulk viscosity can drive acceleration without requiring a cosmological constant or scalar field. Several studies have proposed an inflationary epoch driven by bulk viscous pressure \cite{Diosi/1984,Waga/1986,Barrow/1986,Barrow/1988}, all of which have investigated the role of bulk viscosity in the early universe. In a homogeneous and isotropic universe, a sufficiently large bulk viscosity can lead to negative effective pressure. This characteristic has been suggested as a possible explanation for the late-time acceleration of the universe. The relationship between DE and bulk viscosity in the cosmic medium was first explored in \cite{Zimdahl/2001}. Recent studies have explored the use of viscous fluids as potential candidates for various cosmological roles, such as DM \cite{Velten/2012}, DE \cite{Setare/2010,Cataldo/2005,Brevik/2005,Jean/2011,Brevik/2001,gron/1990,C.E./1940}, or unified models where a single substance serves as both DM and DE concurrently \cite{Li/2009,Hipolito/2009,Montiel/2011}. It has been demonstrated that with a suitable viscosity coefficient, it is possible to achieve an accelerating cosmology without requiring a cosmological constant \cite{Hu/2006,Ren/2006,Gagnon/2011}. This study aims to explain the current acceleration of the universe by considering the bulk viscous pressure within the cosmic fluid, without the need for a DE component in modified $f(T)$ gravity theory.

In this study, we investigate the effects of varying bulk viscosity coefficients on cosmic evolution within the framework of $f(T)$ teleparallel gravity. The model incorporates a perfect fluid with a bulk viscosity given by $\zeta(t)=\zeta_{0}+\zeta_{1}H$, where $\zeta_{0}$ and $\zeta_{1}$ are constants, and $H$ is the Hubble parameter. The exact solutions to the field equations are derived for both cases: (i) $\zeta_{1} \neq0$ and (ii) $\zeta_{1} =0$, assuming the simplest linear form of $f(T) = \alpha T$, where $\alpha \neq 0$. By employing the combined dataset of Hubble $H(z)$, $Pantheon^{+}$, and $BAO$, we constrain the model parameters and then analyze several cosmological parameters. The paper is structured as follows: Sec. \ref{sec2} provides a brief overview of the modified $f(T)$ gravity. In Sec. \ref{sec3}, we introduce the cosmological $f(T)$ model and its corresponding field equations incorporating a bulk viscous fluid. Further, we derive the Hubble parameter $H$ as a function of redshift $z$ for both cases: (i) $\zeta_{1} \neq0$ and (ii) $\zeta_{1} =0$. In Sec. \ref{sec4}, we determine the best-fit values of the viscosity coefficients and the model parameter by analyzing the combined $H(z)+Pantheon^{+}+BAO$ dataset. In addition, we investigate various cosmological aspects, including the deceleration parameter, jerk parameter, energy density, effective pressure, effective EoS parameter, statefinder parameters, and the $Om(z)$ diagnostic, for our $f(T)$ model dominated by bulk viscous matter in Secs. \ref{sec5}, \ref{sec6}, and \ref{sec7}, respectively. In Sec. \ref{sec8}, we provide a summary of our findings.

\section{Brief review of modified $f(T)$ gravity} \label{sec2}

The theory of gravity known as $f(T)$ theory (which generalizes teleparallel gravity) is defined uniquely by the tetrad field \cite{Aldrovandi/2013}. This field consists of an orthonormal set of four-vector fields defined within the tangent space at every point of the Lorentzian manifold. The relationship between the metric and tetrad fields is given by:
\begin{equation}
    ds^2=g_{\mu \nu} dx^{\mu} dx^{\nu}=\eta_{ij} \theta^{i} \theta^{j}.
\end{equation}

Here, we define the following components:
\begin{equation}
    dx^{\mu}=e_{i}^{\,\, \mu}\theta^{i}, \quad \theta^{i}=e^{i}_{\,\, \mu}dx^{\mu},
\end{equation}
where $\eta_{ij} = \text{diag}(1, -1, -1, -1)$  is the metric tensor for Minkowskian spacetime, and $\{e^{i}_{\,\,\mu}\}$ are the components of the tetrad that satisfy the following conditions:
\begin{equation}
e_{i}\,\,^{\mu} e^{i}\,\,_{\nu} = \delta^\mu_\nu, \quad e_{\mu}\,\,^{i} e^{\mu}\,\,_{j} = \delta^i_j. \label{eq:identity}
\end{equation}

In this theory, we use the connection that follows the prescription of Weitzenb\"ock \cite{Aldrovandi/2013},
\begin{equation}
\Gamma^{\alpha}_{\mu \nu}= e_{i}^{\,\, \alpha} \partial_{\mu} e^{i}_{\,\,\nu}=- e^{i}_{\,\,\mu} \partial_{\nu} e_{i}^{\,\, \alpha}.\label{eq:connection}
\end{equation}

With this connection, the components of the torsion tensor are expressed as
\begin{equation}
T^{\alpha}_{\,\,\mu \nu}= \Gamma^{\alpha}_{\,\,\nu \mu}- \Gamma^{\,\,\alpha}_{\mu \nu}= e_{i}^{\,\,\alpha}\left(\partial_{\mu} e^{i}_{\,\,\nu}-\partial_{\nu} e^{i}_{\,\,\mu} \right).
\label{eq:torsion}
\end{equation}

This tensor contributes to defining the contorsion tensor as
\begin{equation}
K^{\mu \nu}_{\,\, \alpha} = -\frac{1}{2}\left(T^{\mu \nu}_{\,\, \alpha} - T^{\nu \mu}_{\alpha} - T_{\alpha}^{\,\, \mu \nu}   \right),
\label{eq:contorsion}
\end{equation}

These objects, torsion, and contorsion combine to form the tensor $S_{\alpha}^{\,\, \mu \nu}$, represented as 
\begin{equation}
    S_{\alpha}^{\,\, \mu \nu} = \frac{1}{2}\left(K^{\mu \nu}_{\,\, \alpha} + \delta^{\mu}_{\alpha} T^{\lambda \mu}_{\,\, \lambda}- \delta^{\nu}_{\alpha} T^{\lambda \mu}_{\,\, \lambda}   \right).
    \label{eq:tensorS}
\end{equation}

The torsion scalar, denoted as $T$, is a fundamental quantity in $f(T)$ gravity, defined as the contraction of the tensor $S_{\alpha}^{\,\, \mu \nu}$ with the torsion tensor $T_{\mu\nu}^{\ \ \ \alpha}$,
\begin{equation}
T = S_{\alpha}^{\,\, \mu \nu} T^{\alpha}_{\,\,\mu \nu}= \frac{1}{2} T^{\alpha \mu\nu} T_{\alpha \mu\nu} + \frac{1}{2} T^{\alpha \mu\nu} T_{ \nu\mu \alpha} - T_{\alpha \mu}^{\,\, \,\, \alpha} T^{\nu \mu}_{\,\,\,\, \nu} .\label{eq:torsion_scalar}
\end{equation}

The gravitational interactions in modified teleparallel geometry, known as $f(T)$ gravity, are described by the following action:
\begin{equation}
\label{7}
    S= \int{\frac{1}{2\kappa^2}f(T)ed^4x} + \int{L_m e d^4x},
\end{equation}
where $\kappa^2 = 8\pi G$, $e=det(e^{i}_{\,\,\mu})$, $f(T)$ is an arbitrary function of the torsion scalar $T$, and $L_m$ is the Lagrangian density for matter fields. When $f(T) = T$, this corresponds to the TEGR.

The gravitational field equation, derived by varying the action (\ref{7}) with respect to the tetrads, is presented below:
\begin{align} \label{1d}
& S_{\mu}^{\,\, \nu \rho} \partial_{\rho}Tf_{TT} + \left[ e^{-1}e_{\mu}^{i}\partial_{\rho}\left(ee_{i}^{\,\, \mu}S_{\alpha}^{\,\, \nu \lambda}\right) + T_{\,\, \lambda \mu}^{\alpha}S_{\alpha}^{\,\, \nu \lambda}\right] f_{T} \nonumber \\
& + \frac{1}{4} \delta_{\mu}^{\nu}f = \frac{\kappa^{2}}{2}\mathcal{T}_{\mu}^{\nu},
\end{align}
where $f_T={\partial f}/{\partial T} $, $ f_{T T}={\partial^2 f}/{\partial T^2}$, and $\mathcal{T}_{\mu}^{\nu}$ is the energy-momentum tensor for the cosmic matter content, expressed as
\begin{equation}\label{2o}
\mathcal{T}_{\mu\nu} = \frac{-2}{\sqrt{-g}} \frac{\delta(\sqrt{-g}L_m)}{\delta g^{\mu\nu}}
\end{equation}

\section{Cosmological $f(T)$ Model} \label{sec3}

To investigate the cosmological implications, we assume the cosmological principle of homogeneity and isotropy of the universe by adopting the flat Friedmann-Lema\^itre-Robertson-Walker (FLRW) metric, which is given by \cite{Ryden}
\begin{equation}
\label{9}
 ds^2= -dt^2 + a^2(t)[dx^2+dy^2+dz^2], 
\end{equation}
where $a(t)$ denotes the scale factor of the universe, it describes the relative size of the universe at a given time $t$. Now, the torsion scalar can be obtained by taking the trace of the non-metricity tensor for the line element (\ref{9}) as
\begin{equation}
T=-6 H^2.
\end{equation}

Here, $H = \dot{a}/a$ denotes the Hubble parameter, which characterizes the rate of expansion of the universe. Furthermore, we consider a universe with bulk viscous matter, where the energy-momentum tensor corresponding to the line element (\ref{9}) is given by
\begin{equation}
\mathcal{T}_{\mu\nu}=(\rho+p_{eff})u_\mu u_\nu + p_{eff}g_{\mu\nu},
\end{equation}
where $\rho$ is the energy density, $p_{eff}$ is the effective pressure, $u^\mu=(1,0,0,0)$ are components of the four velocities of the cosmic fluid, and $g_{\mu\nu}$ is the metric tensor. 

From a fluid dynamics perspective, two viscosity coefficients are commonly discussed in the literature: the bulk viscosity coefficient $\zeta$ and the shear viscosity coefficient $\eta$. In the context of a spatially isotropic universe, which is supported by observations, shear viscosity is often negligible. When a system deviates from thermal equilibrium, an effective pressure is generated to restore equilibrium. In a cosmic fluid, bulk viscosity arises from this effective pressure, expressed as \cite{Brevik/2005,Jean/2011,Brevik/2001,gron/1990,C.E./1940}
\begin{equation} \label{p_eff}
    p_{eff}=p-3\zeta(t) H,
\end{equation}
where $p$ denotes the standard pressure, which is zero for the non-relativistic matter. The term $-3\zeta(t) H$ corresponds to the bulk viscous pressure, where the coefficient of bulk viscosity, $\zeta(t)$, may generally depend on the velocity and acceleration of expansion. In this paper, we investigate a time-dependent bulk viscosity given by \cite{Meng/2009,Meng/2007,Avelino/2010}
\begin{equation}
\zeta(t)=\zeta _{0}+\zeta _{1}\left( \frac{\dot{a}}{a}\right)=\zeta_0+\zeta_1 H,
\end{equation}
where $\zeta_0$ and $\zeta_1$ are bulk viscous parameters. The time-dependent bulk viscosity we consider is a linear combination of two terms: the first is a constant, and the second is proportional to the Hubble parameter, indicating the dependence of viscosity on speed. 

The corresponding modified Friedmann equations, describing the universe dominated by bulk viscous matter in $f(T)$ gravity, are given by \cite{Sharif/2013}
\begin{equation}
3H^{2}=\frac{1}{2f_{T}}\left( \kappa^2 \rho -\frac{f}{2}\right)  \label{F1}
\end{equation}%
and 
\begin{equation}
\dot{H}+3H^{2}+\frac{\dot{f_{T}}}{f_{T}}H=-\frac{1}{2f_{T}}\left( \kappa^2 p_{eff}+
\frac{f}{2}\right) \text{.}  \label{F2}
\end{equation}%

In this paper, we assume the following $f(T)$ function to investigate the dynamics of a universe with viscosity, given by
\begin{equation}
    f(T) = \alpha T, \quad \alpha \neq 0.
\end{equation}

For this particular functional form, with $\kappa^2=1$, the modified Friedmann equations (\ref{F1}) and (\ref{F2}) describing the universe dominated by the bulk viscous matter are given by
\begin{equation}
\rho =3\alpha H^{2},  \label{3h}
\end{equation}%
and 
\begin{equation}
p_{eff}=-\alpha (2 \dot{H}+3 H^{2}).  \label{3i}
\end{equation}%

Specifically, for the case $\alpha=1$, one can recover the standard Friedmann equations of GR. Here, we consider that non-relativistic matter dominates the universe i.e. $p=0$. From Eq. (\ref{p_eff}) and the modified Friedmann equation (\ref{3i}), we have
\begin{equation}
\overset{.}{H}+\frac{3\left( \alpha -\zeta _{1}\right) }{2\alpha }H^{2}-%
\frac{3\zeta _{0}}{2\alpha }H=0.
\end{equation}

By employing the relationship $ \frac{1}{H} \frac{d}{dt}= \frac{d}{dln(a)}$ (where $a=\frac{1}{1+z}$), we can transform the given equation into a first-order differential equation for the Hubble parameter as,
\begin{equation}\label{20}
\frac{dH}{d\ln \left( a\right) }+\frac{3\left( \alpha -\zeta _{1}\right) }{2\alpha }H-\frac{3\zeta _{0}}{2\alpha }=0.
\end{equation}

Integrating Eq. (\ref{20}), we derive the expression for the Hubble parameter in terms of redshift as follows:
\begin{equation} \label{Hz}
H\left( z\right) =H_{0}\left( 1+z\right) ^{\frac{3}{2}}+\frac{\zeta _{0}}{%
\left( \alpha -\zeta _{1}\right) }\left\{ 1-\left( 1+z\right) ^{\frac{%
3\left( \alpha -\zeta _{1}\right) }{2\alpha }}\right\}.   
\end{equation}

Here, $H(z=0) = H_0$ denotes the current value of the Hubble parameter.  Specifically, for the scenario $\alpha = 1$, with $\zeta_{0} = 0$ and $\zeta_{1} = 0$, the solution simplifies to $H(z) = H_0 (1 + z)^{\frac{3}{2}}$, which describes the non-viscous matter-dominated universe. Now, we consider the following two different cases concerning the viscous parameters, as commonly discussed in the literature:

\textbf{Case I}: The viscosity coefficient depends on velocity, where both $\zeta_0$ and $\zeta_1$ are non-zero.

\textbf{Case II}: The viscosity coefficient does not depend on velocity, where $\zeta_1 = 0$ and $\zeta_0$ is non-zero. In this scenario, the expression for the Hubble parameter can be expressed as
\begin{equation}
H\left( z\right) =H_{0}\left( 1+z\right) ^{\frac{3}{2}}+\frac{\zeta _{0}}{ \alpha }\left\{ 1-\left( 1+z\right) ^{\frac{3}{2}}\right\}.     
\end{equation}

In Eq. (\ref{Hz}), we present our model in which key cosmological aspects are defined by the model parameters ($H_0$, $\alpha$, $\zeta_0$, $\zeta_1$). In the following part, we investigate the behavior of cosmological parameters by constraining these model parameters using current observational datasets.

\section{Observational data} \label{sec4}

In this section, we assess the consistency of our $f(T)$ model dominated by bulk viscous matter by verifying its agreement with recent observational data. In our analysis, we include various observational datasets, such as the $H(z)$ dataset, the $Pantheon^{+}$ sample of the SNe Ia dataset, and the $BAO$ dataset. Further, we use the \textit{emcee} package \cite{Mackey_2013} in Python, which employs the Monte Carlo Markov Chain (MCMC) technique. 
This approach enabled us to constrain the model parameters ($H_0$, $\alpha$, $\zeta_0$, $\zeta_1$), facilitating an exploration of the posterior distribution of the parameter space across all both distinct cases i.e. $\zeta_1 \neq 0$ (Case I) and $\zeta_1=0$ (Case II). Here, we describe the observational data that we have utilized:

\begin{itemize}
\item \textbf{$H(z)$ data}: The Hubble dataset provides a valuable means to directly constrain the Hubble rate $H(z)$ at different redshifts. Here, we use 31 data points collected from studies by \cite{Jimenez_2003,Simon_2005,Stern_2010,Moresco_2012,Zhang_2014,Moresco_2015,Moresco_2016}. The $H(z)$ method involves using spectroscopic dating techniques on galaxies that evolve passively to estimate the age difference between two galaxies at different redshifts. This age difference enables the inference of $\frac{dz}{dt}$ from observations, which allows for the computation of $H(z) = -\frac{1}{1+z} \frac{dz}{dt}$. Hence, $H(z)$ data are regarded as highly reliable because they are independent of any particular cosmological model, do not necessitate intricate integration, and depend on the absolute age determination of galaxies \cite{Jimenez_2002}.

\item \textbf{$Pantheon^{+}$ data}: Recent observational discoveries related to SNe Ia have corroborated the existence of the accelerated expansion phase of the universe. In the previous 20 years, there has been a significant accumulation of data from SNe Ia samples. Here, we use the $Pantheon+$ sample \cite{Scolnic_2022,Brout_2022}, which is one of the most comprehensive compilations of SNe Ia data. It includes 1701 light curves of 1550 SNe Ia within the redshift range of $[0.001, 2.26]$.

\item \textbf{$BAO$ data}: The $BAO$ studies the oscillations that arose in the early universe as a result of cosmological perturbations in the fluid containing photons, baryons, and DM. This fluid was tightly coupled due to Thomson scattering. Here, we incorporate $BAO$ measurements that include data from the Sloan Digital Sky Survey (SDSS), the Six Degree Field Galaxy Survey (6dFGS), and the Baryon Oscillation Spectroscopic Survey (BOSS) \cite{BAO1, BAO2, BAO3, BAO4, BAO5, BAO6}.

\end{itemize}

To find the fitting outcomes in our MCMC study, we use 100 walkers and 1000 steps, and we incorporate the following priors: $H_0 \in [60,80] , \text{km/s/Mpc}, \quad \alpha \in [-2,2], \quad \zeta_{0} \in [0,100], \quad \zeta_{1} \in [-2,2]$. Furthermore, for the combined $H(z)+Pantheon^{+}+BAO$ dataset, we apply the following likelihood and Chi-square calculations ($\mathcal{L} \propto exp(-\chi^2/2)$) as
\begin{eqnarray}
\mathcal{L}_{joint} &=& \mathcal{L}_{H(z)} \times \mathcal{L}_{Pan^{+}} \times \mathcal{L}_{BAO},\\
\chi^{2}_{joint} &=& \chi^{2}_{H(z)} + \chi^{2}_{Pan^{+}}+ \chi^{2}_{BAO}.
\end{eqnarray}
where
\begin{eqnarray}
\chi^{2}_{H(z)} &=& \sum_{i=1}^{31} \frac{\left[H(\theta_{s}, z_{i})-
H_{obs}(z_{i})\right]^2}{\sigma(z_{i})^2},\\
\chi^{2}_{Pan^{+}} &=& \sum_{i,j=1} ^{1701} \Delta \mu_{i} \left(
C_{Pan^{+}}^{-1}\right)_{ij} \Delta \mu_{j},\\
\chi _{BAO}^{2} &=& X^{T}C_{BAO}^{-1}X\,.
\end{eqnarray}

For $\chi^{2}_{H(z)}$, $H(\theta_{s}, z_{i})$ denotes the model-predicted Hubble parameter at redshift $z_{i}$, determined by the model parameters $\theta_{s}=(H_0, \alpha, \zeta_0, \zeta_1)$. $H_{obs}(z_{i})$ denotes the observed Hubble parameter at redshift $z_{i}$, and $\sigma(z_{i})$ is the standard error associated with the observed value at that redshift. For $\chi^{2}_{Pan^{+}}$, $\Delta \mu_{i}=\mu_{\rm th}-\mu_{\rm obs}$ denotes the difference between the distance modulus of the $i_{th}$ SNe Ia data point and the corresponding theoretical prediction, while $C_{Pan^{+}}^{-1}$ is the inverse covariance matrix of the $Pan^{+}$ sample. Further, the calculated theoretical value of the distance modulus is defined as $\mu _{th}=5log_{10}\frac{d_{L}(z)}{1Mpc}+25$, where $d_{L}(z)=c(1+z)\int_{0}^{z}\frac{dy}{H(y,\theta_{s} )}$ is the luminosity distance \cite{Planck/2018}. For $\chi^{2}_{BAO}$, $X$ is a vector that changes based on the specific survey under consideration, and $C_{BAO}^{-1}$ is the inverse covariance matrix for the BAO data \cite{BAO6}.

Figs. \ref{F_C1} and \ref{F_C2} display the $1-\sigma$ and $2-\sigma$ likelihood contours for the model parameters $H_0$, $\alpha$, $\zeta_0$, and $\zeta_1$ for both cases using the combined $H(z)+Pantheon^{+}+BAO$ dataset. For Case I ($\zeta_1 \neq 0$), the best-fit values of the model parameters are $H_0=60.0^{+2.0}_{-1.9}$ $\text{km/s/Mpc}$, $\alpha=1.01^{+0.10}_{-0.098}$, $\zeta_0=40.1^{+1.9}_{-2.0}$ $\text{kg/m} \cdot \text{s}^{-1}$, and $\zeta_1=0.123^{+0.093}_{-0.088}$ $\text{kg/m}$. For Case II ($\zeta_1 = 0$), the best-fit values are $H_0=67.5^{+1.3}_{-1.3}$ km/s/Mpc, $\alpha=0.94^{+0.14}_{-0.13}$, and $\zeta_0=34.7^{+2.0}_{-2.0}$ $\text{kg/m} \cdot \text{s}^{-1}$. Recent observational data suggests that the Hubble parameter $H$ is decreasing over time as the universe expands. Here, we investigate the evolution of the Hubble parameter $H(z)$ versus redshift $z$, for both cases using the constrained values of the model parameters, as depicted in Fig. \ref{F_H}. The figure indicates a decrease in the value of the Hubble parameter as the universe evolves, consistent with observational data. In particular, we observe that $H$ increases as $z$ increases for both cases. According to recent observational data from Planck collaborators \cite{Planck/2018}, the Hubble constant has been determined to be $H_0 = 67.4 \pm 0.5$ km s$^{-1}$ Mpc$^{-1}$. This value is consistent with the Hubble constant for Case II, while Case I shows a higher deviation. The lower value of $H_0$ in Case I compared to Case II can be attributed to several factors. First, the interplay between $\zeta_0$ and $\zeta_1$ imposes additional constraints on the model parameters, further influencing the value of $H_0$. In addition, the combined $H(z) + Pantheon^{+} + BAO$ dataset, used to constrain the parameters, favors a lower $H_0$ in Case I due to the increased viscous effects during the recent cosmic expansion.

\begin{widetext}

\begin{figure}[h]
\centering
\includegraphics[scale=0.7]{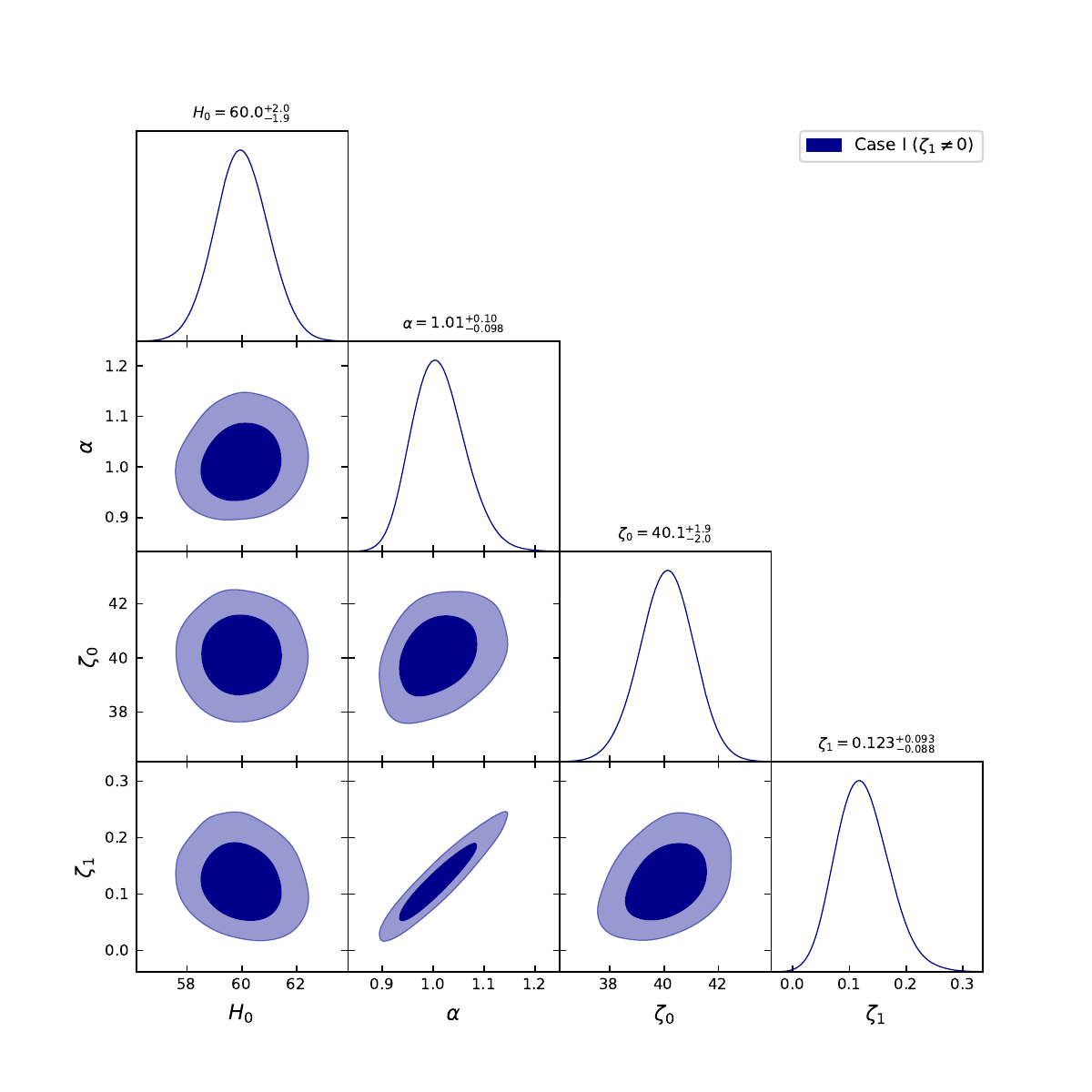}
\caption{The constrained values for the model parameters $H_0$, $\alpha$, $\zeta_0$, and $\zeta_1$ using the combined $H(z)+Pantheon^{+}+BAO$ dataset are shown (Case I). The regions colored in dark blue indicate the $1-\sigma$ confidence level (CL), while the regions shaded in light blue indicate the $2-\sigma$ CL.}
\label{F_C1}
\end{figure}

\begin{figure}[h]
\centering
\includegraphics[scale=0.6]{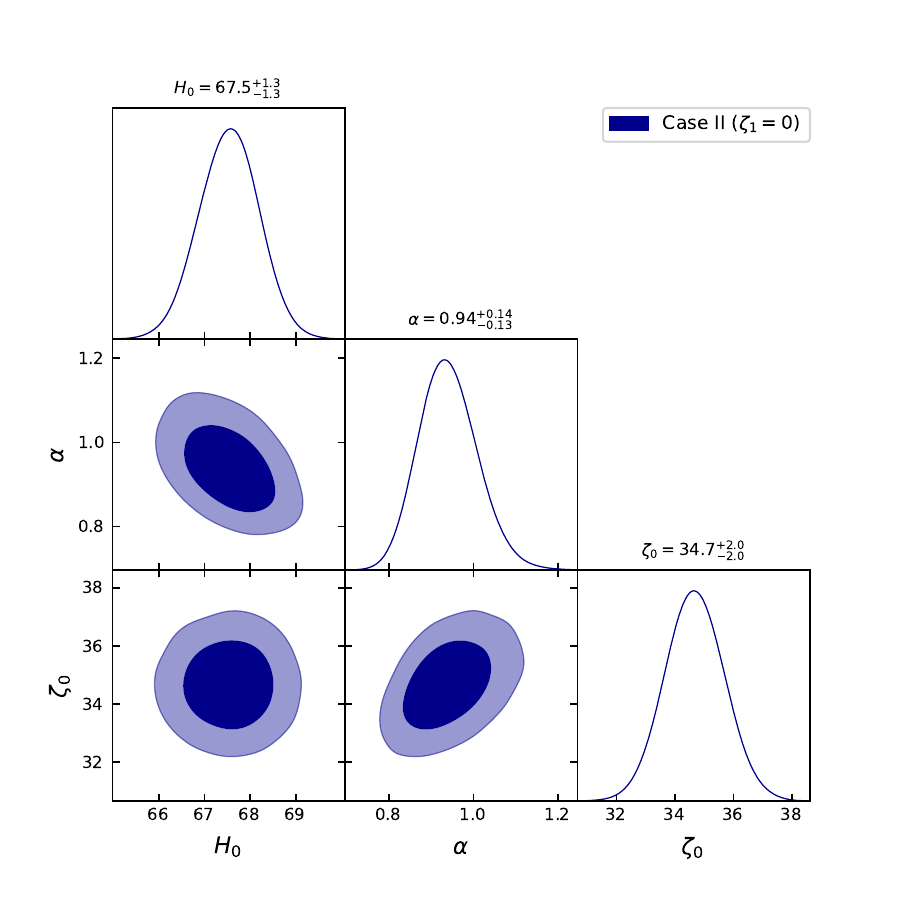}
\caption{The constrained values for the model parameters $H_0$, $\alpha$, and $\zeta_0$ using the combined $H(z)+Pantheon^{+}+BAO$ dataset are shown (Case II). The regions colored in dark blue indicate the $1-\sigma$ confidence level (CL), while the regions shaded in light blue indicate the $2-\sigma$ CL.}
\label{F_C2}
\end{figure}

\begin{figure}[h]
   \begin{minipage}{0.48\textwidth}
     \centering
     \includegraphics[width=\linewidth]{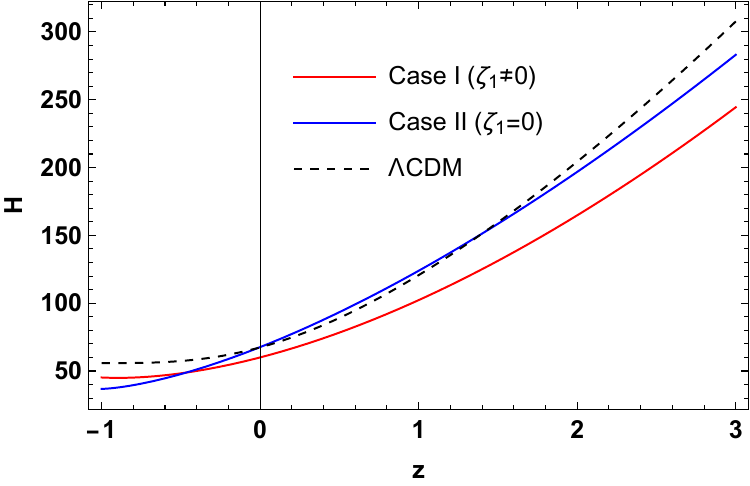}
     \caption{The behavior of the Hubble parameter $H$ versus redshift $z$: A comparison between the viscous model and $\Lambda$CDM, showing consistency with Planck collaboration values ($H_0 = 67.4 \pm 0.5$ and $\Omega_{m0} = 0.315 \pm 0.007$) \cite{Planck/2018}.}\label{F_H}
   \end{minipage}\hfill
   \begin{minipage}{0.48\textwidth}
     \centering
     \includegraphics[width=\linewidth]{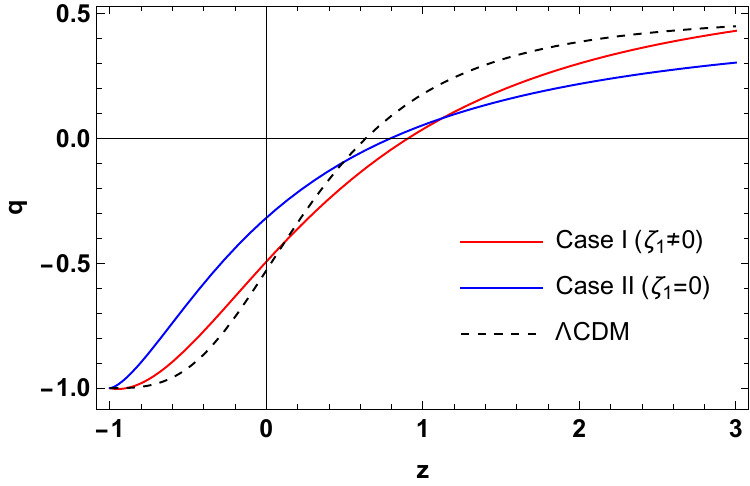}
     \caption{The behavior of the deceleration parameter $q$ versus redshift $z$: Transition from deceleration to acceleration in viscous models and $\Lambda$CDM ($H_0 = 67.4 \pm 0.5$ and $\Omega_{m0} = 0.315 \pm 0.007$) \cite{Planck/2018}.}\label{F_q}
   \end{minipage}
\end{figure}

\begin{figure}[h]
   \begin{minipage}{0.48\textwidth}
     \centering
     \includegraphics[width=\linewidth]{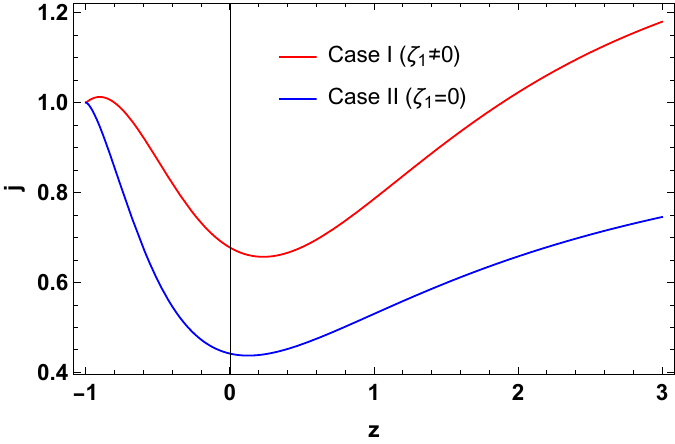}
     \caption{The behavior of the jerk parameter $j$ versus redshift $z$.}\label{F_j}
   \end{minipage}\hfill
   \begin{minipage}{0.48\textwidth}
     \centering
     \includegraphics[width=\linewidth]{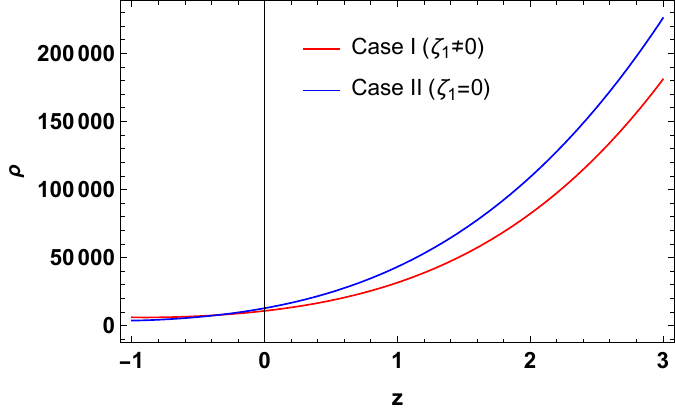}
     \caption{The behavior of the energy density $\rho$ versus redshift $z$.}\label{F_rho}
   \end{minipage}
\end{figure}

\begin{figure}[h]
   \begin{minipage}{0.48\textwidth}
     \centering
     \includegraphics[width=\linewidth]{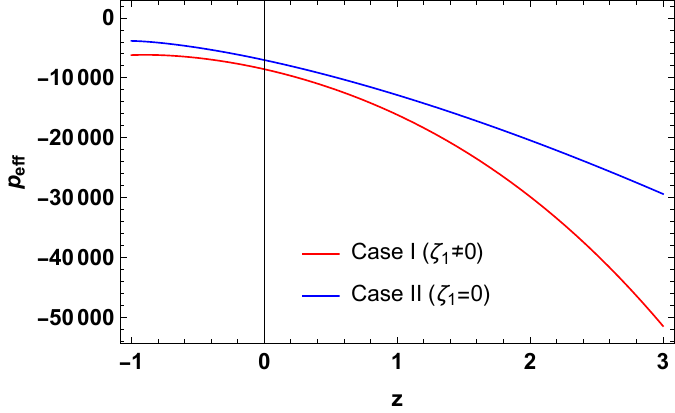}
     \caption{The behavior of the effective pressure $p_{eff}$ versus redshift $z$.}\label{F_p}
   \end{minipage}\hfill
   \begin{minipage}{0.48\textwidth}
     \centering
     \includegraphics[width=\linewidth]{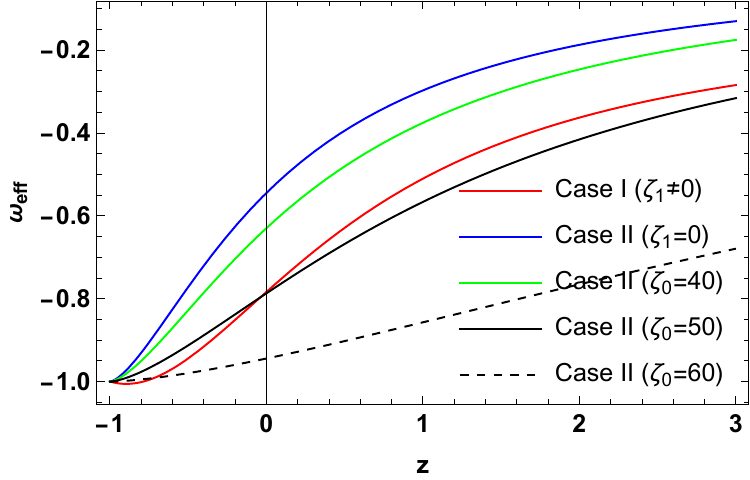}
     \caption{The behavior of the EoS parameter $\omega_{eff}$ versus redshift $z$.}\label{F_w}
   \end{minipage}
\end{figure}

\end{widetext}

\section{Cosmological applications} \label{sec5}

In this section, we will explore several cosmological applications of our $f(T)$ model dominated by bulk viscous matter. Specifically, we will analyze important cosmological parameters, including the deceleration parameter, jerk parameter, and equation of state (EoS) parameter.

\subsection{The deceleration parameter}

The deceleration parameter $q$ is a crucial indicator of the universe's expansion dynamics. Its sign determines whether the expansion is accelerating or decelerating. In our model, the behavior of $q$ is particularly insightful. For $q>0$, the universe experiences decelerating expansion, suggesting a dominance of gravitational attraction. A value of $q=0$ implies a constant rate of expansion, while for $-1<q<0$, the expansion accelerates, indicating a possible influence of DE. Moreover, our model predicts intriguing scenarios for extreme values of $q$. When $q=-1$, the universe undergoes exponential expansion, known as de Sitter (dS) expansion. This phase is associated with a cosmological constant or vacuum energy dominating the universe's dynamics. For $q<-1$, the expansion becomes super-exponential, suggesting a rapid and extreme acceleration of the universe's size. The deceleration parameter is defined as
\begin{equation}
q= -\dfrac{1}{aH^2}\dfrac{d^2a}{dt^2}= -1+\frac{(1+z)}{H}\frac{dH}{dz}. 
\end{equation}

Fig. \ref{F_q} illustrates a notable transition in the deceleration parameter, marking the shift from a decelerated phase ($q > 0$) to an accelerated phase ($q < 0$) of the universe's expansion. This transition occurs for the constrained values of the model parameters. Specifically, the transition redshifts are found to be $z_{tr} \approx 0.90$ and $z_{tr} \approx 0.80$, for the respective cases. In addition, the present-day values of the deceleration parameter are determined to be $q_{0} \approx -0.49$ and $q_{0} \approx -0.32$ for the respective cases \cite{Cunha1,Cunha2,Nair,Koussour1,Koussour2,Koussour3,DP1, DP2,DP3,DP4,DP5,DP6,DP7,DP8}.

\subsection{The jerk parameter}

In the context of cosmology, the jerk parameter $j$ is a dimensionless quantity that describes the rate at which the deceleration of the universe's expansion is changing. More precisely, it quantifies the rate of change of the third derivative of the scale factor $a(t)$ with respect to time $t$. Positive values of $j$ imply an acceleration in the rate of cosmic acceleration, whereas negative values indicate a deceleration in this rate. The jerk parameter is defined as \cite{Visser1,Visser2}
\begin{equation}
\label{jerk}
    j= \dfrac{1}{aH^3}\dfrac{d^3a}{dt^3}=q(2q+1)+(1+z)\frac{dq}{dz}. 
\end{equation}

The jerk parameter is frequently employed to distinguish between different DE scenarios. In Fig. \ref{F_j}, we display the behavior of $j(z)$ for both cases ($\zeta_{1} \neq0$ and $\zeta_{1} =0$). The present-day values of the jerk parameter are approximately $j_{0} \approx 0.68$ and $j_{0} \approx 0.44$ for the respective cases. These values indicate the deviation of $j$ from the flat $\Lambda$CDM model ($j=1$) for both cases, as constrained by the model parameters. Thus, considering our $f(T)$ model dominated by bulk viscous matter, characterized by $j_0 > 0$ and $q_0 < 0$, it becomes apparent that the dynamic DE model under scrutiny stands out as the most plausible explanation for the current acceleration of the universe.

\subsection{The EoS parameter}

The EoS parameter $\omega$ is instrumental in categorizing the different epochs of accelerated and decelerated expansion in the universe \cite{Myrzakulov/2023}. Different values of $\omega$ correspond to different types of cosmic fluids dominating the universe's dynamics: (a) $\omega=1$ represents a stiff fluid, which has an energy density that decreases faster than its pressure as the universe expands; (b) $\omega=\frac{1}{3}$ depicts the radiation-dominated phase, where radiation is the dominant component, characterized by a fast decrease in energy density with expansion; (c) $\omega=0$ indicates the matter-dominated phase, where matter is the dominant component, with an energy density that decreases more slowly with expansion compared to radiation. For more exotic scenarios, such as quintessence and phantom DE, the EoS parameter takes on values less than -1 and greater than -1, respectively. Quintessence refers to a hypothetical form of DE with an EoS less than -1/3 but greater than -1, leading to an accelerating expansion of the universe \cite{quint}. On the other hand, phantom energy has an EoS less than -1, causing a rapid expansion known as the Big Rip scenario, where the universe eventually tears apart due to the ever-increasing expansion rate \cite{phant}. In the standard $\Lambda$CDM model, the EoS parameter for the cosmological constant $\Lambda$ is exactly -1, representing a constant energy density throughout the universe's evolution. This model has been remarkably successful in explaining various cosmological observations, including the accelerated expansion of the universe. The effective EoS parameter is defined as 
\begin{equation}
\omega_{eff}=\frac{p_{eff}}{\rho}=-\frac{\zeta_{0}+\zeta_{1} H}{\alpha  H}.    
\end{equation}

In this context, it is necessary to investigate the behavior of energy density and pressure components in the presence of bulk viscosity. From Figs. \ref{F_rho} and \ref{F_p}, it is evident that the energy density $\rho$ behaves as expected, showing a positive trend and decreasing as the universe expands, eventually approaching zero in the distant future. On the other hand, the effective pressure $p_{eff}$ demonstrates a negative trend for both cases ($\zeta_{1} \neq0$ and $\zeta_{1} =0$). The negative pressure observed confirms the existence of a mysterious DE component. This characteristic of negative pressure in bulk viscosity makes it a plausible candidate for driving cosmic acceleration.

For both cases ($\zeta_{1} \neq0$ and $\zeta_{1} =0$), the behavior in Fig. \ref{F_w}, depicting the effective EoS for the cosmic viscous fluid, aligns with the recently observed acceleration of the universe. It behaves like quintessence DE ($-1<\omega<-\frac{1}{3}$) in the present epoch and converges to the EoS of $\Lambda$CDM ($\omega_{\Lambda}=-1$) in the distant future. Further, the present-day values of the jerk parameter are approximately $\omega_{0} \approx -0.78$ and $\omega_{0} \approx -0.55$ for the respective cases \cite{EoS1,EoS2,EoS3,EoS4}. The same figure shows the evolution of $\omega_{eff}$ for different cases of the bulk viscosity parameter $\zeta_0$. Specifically, we observe a significant decrease in $\omega_{eff}$, which approaches $-1$ as $\zeta_0$ increases, indicative of a cosmological constant-like behavior.

\section{Statefinder diagnostic} \label{sec6}

As numerous DE models emerge, distinguishing between them qualitatively or quantitatively becomes essential. To tackle this challenge, V. Sahni et al. \cite{Sahni1} introduced novel geometric diagnostic parameters, referred to as statefinder parameters $(r, s)$. The
statefinder parameters are defined as \cite{Sahni1,Sahni2}
\begin{equation}
r=\frac{\overset{...}{a}}{aH^{3}},\text{ \ \ }s=\frac{r-1}{3\left( q-\frac{1%
}{2}\right) }.  
\end{equation}

\begin{figure}[h]
\includegraphics[scale=0.7]{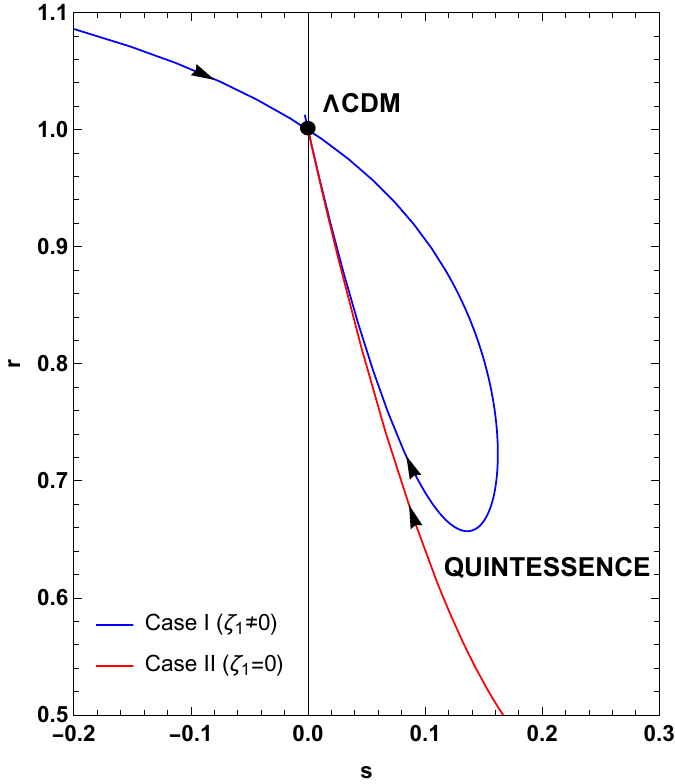}
\caption{The behavior of the $(r, s)$ plane for $z \in (-1,3)$.}
\label{F_rs}
\end{figure}

\begin{figure}[h]
\includegraphics[scale=0.7]{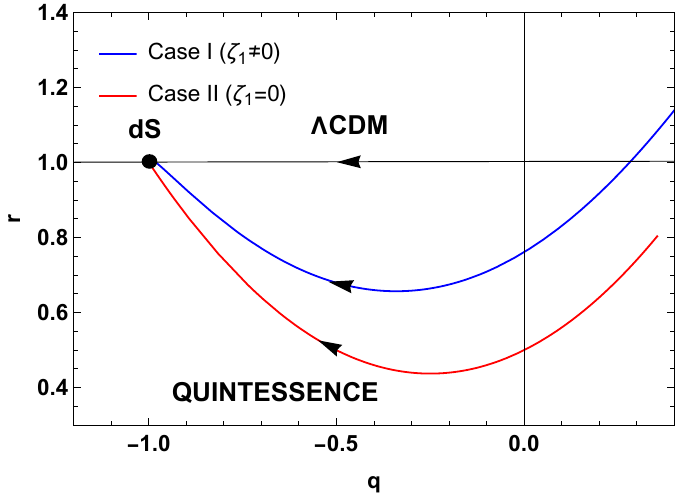}
\caption{The behavior of the $(r, q)$ plane for $z \in (-1,3)$.}
\label{F_rq}
\end{figure}

For example, in the $\Lambda$CDM model, the statefinder pair is represented as $(r, s) = (1, 0)$. Conversely, models such as the Chaplygin gas model, which exhibits an EoS transitioning from a stiff fluid to a cosmological constant, are denoted by values of $(r, s)$ where $r > 1$ and $s < 0$. Quintessence models, on the other hand, are characterized by $(r, s)$ values where $r < 1$ and $s > 0$. In Figs. (\ref{F_rs}) and (\ref{F_rq}), we depict the behavior of our $f(T)$ model, which is dominated by bulk viscous matter, in the $r$-$s$ and $r$-$q$ planes for both cases ($\zeta_{1} \neq0$ and $\zeta_{1} =0$). The trajectories in both planes consistently fall within the region where $r < 1$, $s > 0$, and $q < 0$, indicating that our $f(T)$ model follows the quintessence scenario. Additionally, we observe that for both cases, the trajectories in the $r$-$s$ plane converge to the $\Lambda$CDM fixed point, while those in the $r$-$q$ plane converge to a dS phase $(r, q) = (1, -1)$ in the far future.

\section{$Om(z)$ diagnostics} \label{sec7}

The $Om(z)$ diagnostic is a recently proposed method that provides an effective means of distinguishing between various DE models \cite{Sahni3}. It is simpler compared to the statefinder analysis, as it involves a formula that includes only the Hubble parameter. For the spatially flat universe, the formula is given by:
\begin{equation}
Om\left( z\right) =\frac{\left( \frac{H\left( z\right) }{H_{0}}\right) ^{2}-1%
}{\left( 1+z\right) ^{3}-1},     
\end{equation}

\begin{figure}[h]
\includegraphics[scale=0.7]{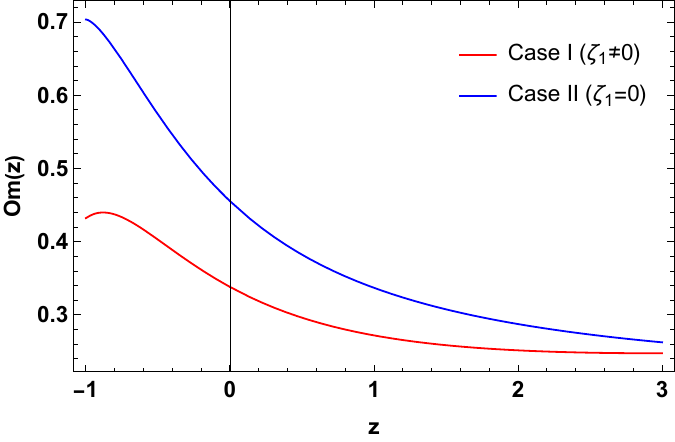}
\caption{The behavior of the $Om(z)$ diagnostic versus redshift $z$.}
\label{F_Om}
\end{figure}

The negative slope of $Om(z)$ corresponds to quintessence-type behavior. In contrast, a positive slope signifies phantom behavior. A constant $Om(z)$ signifies the $\Lambda$CDM model. From Fig. \ref{F_Om}, it is clear that the $Om(z)$ diagnostic in both cases ($\zeta_{1} \neq0$ and $\zeta_{1} =0$) shows a negative slope as the universe expands. We can conclude that the behavior of the $Om(z)$ diagnostic for our $f(T)$ model dominated by bulk viscous matter aligns with the behavior observed for previous cosmological parameters.

\section{Conclusion} \label{sec8}

From a geometric perspective, general relativity can be represented using three equivalent geometric formulations: the curvature representation (where the torsion and non-metricity vanish), the teleparallel representation (where the curvature and non-metricity vanish), and the symmetric teleparallel representation (where the curvature and torsion vanish). In this study, we focused on teleparallel gravity, where the fundamental geometric quantity describing gravitational interactions is the torsion $T$ \cite{Aldrovandi/2013}. Torsion geometrically characterizes the local twisting or rotational aspects of a manifold induced by a connection (or covariant derivative). Over the past two decades, several studies have explored various physical aspects of teleparallel gravity \cite{Paliathanasis/2016,Salako/2013,Myrzakulov/2011, Bamba/2011,Chen/2011,Li/2011,Capozziello/2011,Liu/2012,Cai/2011,Jamil/2013,Rodrigues/2016,Koussour/2022,Nunes/2016}.

Here, we explored the effects of varying bulk viscosity coefficients $\zeta(t)=\zeta_{0}+\zeta_{1}H$ on cosmic evolution within the framework of $f(T)$ teleparallel gravity. From a hydrodynamic viewpoint, introducing viscosity coefficients into the cosmic matter content is natural, as the ideal features of a fluid are fundamentally abstract. We specifically focused on two widely discussed cases of bulk viscosity coefficients: (i) $\zeta_{1} \neq0$ and (ii) $\zeta_{1} =0$. We derived the Hubble parameter $H$ as a function of redshift $z$ for both cases using a linear $f(T)$ model, specifically $f(T) = \alpha T$ where $\alpha \neq 0$. Using the combined $H(z)+Pantheon^{+}+BAO$ dataset, we obtained observational constraints on the model parameter and viscosity coefficients for both cases. In Case I ($\zeta_1 \neq 0$), we found the best-fit values of the model parameters to be $H_0=60.0^{+2.0}_{-1.9}$ $\text{km/s/Mpc}$, $\alpha=1.01^{+0.10}_{-0.098}$, $\zeta_0=40.1^{+1.9}_{-2.0}$ $\text{kg/m} \cdot \text{s}^{-1}$, and $\zeta_1=0.123^{+0.093}_{-0.088}$ $\text{kg/m}$. On the other hand, for Case II ($\zeta_1 = 0$), the best-fit values are $H_0=67.5^{+1.3}_{-1.3}$ km/s/Mpc, $\alpha=0.94^{+0.14}_{-0.13}$, and $\zeta_0=34.7^{+2.0}_{-2.0}$ $\text{kg/m} \cdot \text{s}^{-1}$.

Our analysis revealed a significant transition in the deceleration parameter, indicating a shift from a decelerated phase ($q > 0$) to an accelerated phase ($q < 0$) of the universe's expansion. This transition occurred for constrained values of the model parameters, with transition redshifts at approximately $z_{tr} \approx 0.90$ and $z_{tr} \approx 0.80$ for the respective cases. Furthermore, the present-day values of the deceleration parameter were determined to be $q_{0} \approx -0.49$ and $q_{0} \approx -0.32$ for the respective cases. Recent studies have extensively investigated the impact of viscosity in various frameworks, including $f(R,T)$ gravity and holographic dark energy models \cite{Singh1,Singh2,Singh3}. These analyses consistently reveal a common pattern: a transition from a decelerated phase to an accelerated phase in the universe's expansion. The analysis of the jerk parameter $j(z)$ indicated present-day values of approximately $j_{0} \approx 0.68$ and $j_{0} \approx 0.44$ for the respective cases. These values suggest a deviation of $j$ from the flat $\Lambda$CDM model as constrained by the model parameters. Further investigation into the energy density $\rho$ and effective pressure $p_{eff}$ components in the presence of bulk viscosity confirmed the existence of a mysterious DE component, with negative pressure supporting cosmic acceleration. The behavior of the effective EoS for the cosmic viscous fluid aligned with the observed acceleration of the universe, resembling quintessence DE in the present epoch and converging to the EoS of $\Lambda$CDM in the distant future. The trajectories in both the $r$-$s$ and $r$-$q$ planes consistently align with the quintessence scenario. Moreover, the $Om(z)$ diagnostic exhibited a negative slope, consistent with previous cosmological parameters. These results further support the viability of our $f(T)$ model dominated by bulk viscous matter in explaining cosmic acceleration.

\section*{Acknowledgment}
This research was funded by the Science Committee of the Ministry of Science and Higher Education of the Republic of Kazakhstan (Grant No. AP23487178).

\section*{Data Availability Statement}
There are no new data associated with this article.

\end{document}